\documentclass[prl,nofootinbib,twocolumn,superscriptaddress]{revtex4}

\def\mysection#1{{\bf #1.} }

\def\lsim{\mathrel{\rlap{\lower4pt\hbox{\hskip1pt$\sim$}}
    \raise1pt\hbox{$<$}}}         %less than or approx. symbol
\def\gsim{\mathrel{\rlap{\lower4pt\hbox{\hskip1pt$\sim$}}
    \raise1pt\hbox{$>$}}}         %greater than or approx. symbol
\def\kpirc{k\pi r_c }
\newcommand{\beq}{\begin{eqnarray}}
\newcommand{\eeq}{\end{eqnarray}}
\begin{document}
\begin{titlepage}

%\vskip -.6in
%\flushleft  \vspace*{-1.5cm}{\small BNL-HET-04/5, LBNL-55324}
%\vskip .35in
\begin{center}
{\Large \bf Flavor from Minimal Flavor Violation \& a Viable Randall-Sundrum Model}
\vskip .17in
{\bf A. Liam Fitzpatrick}$^a$,
{\bf Gilad Perez}$^b$ and
{\bf Lisa Randall}$^a$
\vskip .18in
\end{center}
\vskip .04in
$^b $\textit{C.~N.~Yang Institute for Theoretical Physics,
State University of New York, Stony Brook, NY 11794-3840, USA;}
\\
$^{a,b} $\textit{Jefferson Laboratory of Physics, Harvard University
Cambridge, Massachusetts 02138, USA;}
\\
$^b$\textit{Physics Department, Boston University
Boston, Massachusetts 02215, USA}

\begin{center} {\bf Abstract}\\\end{center}
We present a variant of the warped extra dimension, Randall-Sundrum (RS), framework  which 
is based on 
 five dimensional (5D) minimal flavor violation (MFV),
in which
the only sources of flavor breaking are two 5D {\emph{anarchic}}
Yukawa matrices.
The Yukawa
matrices also control the bulk masses, which are responsible for 
the resulting flavor structure and mass hierarchy in the low energy
theory.
An interesting result of this set-up is that at low energies the theory flows to 
next to MFV model where flavor violation is dominantly coming from the third generation.
Low energy flavor violation
 is further suppressed by a single parameter that dials the amount of violation
in the up or down sector. 
There is therefore
a sharp limit in which there is no flavor violation in the down type quark sector
which, remarkably, is favored when we fit for the flavor parameters. 
This mechanism is used to eliminate
the current RS flavor and CP problem 
even
with a Kaluza-Klein scale as low as 2 TeV!
Our construction also suggests that economic supersymmetric and non-supersymmetric, strong dynamic-based, flavor models may
be built based on the same concepts.\vskip .2in

\end{titlepage}
\newpage
\renewcommand{\thepage}{\arabic{page}}
\setcounter{page}{1}

\mysection{Introduction}
The standard model (SM) 
 agrees very well with data. 
However, it is widely perceived to be an incomplete theory.
In particular, in the SM, the hierarchy between the Planck scale and
the electroweak (EW) symmetry breaking (EWSB) scale
is unnatural since the Higgs mass is ultra-violet (UV) sensitive.

Solutions to the hierarchy problem therefore involve extending the
SM at just above 
the EWSB scale which, in
general, 
spoils the good agreement of the SM with data when trying to explain flavor as well.   Given this inherent tension, it is important to
identify new physics (NP) frameworks that preserve the 
approximate symmetries
 of the SM. 

In this letter
 we consider the Randall-Sundrum scenario (RS1) \cite{rs1}, which 
potentially
provides an elegant solution to the hierarchy problem.
In this framework, 
due to warped higher-dimensional spacetime, the
mass scales in an effective $4D$ description
depend on location in an extra dimension:
the Higgs sector is localized at the ``TeV'' brane where it is protected
by a low warped-down fundamental scale of order a TeV while $4D$ gravity
is localized near the ``Planck'' brane which has a Planckian fundamental scale.

In the original RS1 model, the entire SM was localized 
on the TeV brane. In this set-up, 
flavor issues are sensitive to the UV completion of the RS1
effective field theory: there is no
understanding of the hierarchies in fermion masses or of
smallness of flavor changing neutral currents
(FCNCs) from higher-dimensional operators that would be
too large if suppressed only by the warped-down cut-off $\sim$ TeV.
Similar tension
 arises when considering the model predictions regarding EW precision tests.

Allowing the SM fermions and gauge fields to propagate in the
bulk 
gives an opportunity to explain flavor, and
makes flavor issues UV-insensitive as follows. The 
light fermions can be localized near the Planck brane
(using a $5D$ fermion mass parameter \cite{gn, gp}) where the effective cut-off 
is much higher than
TeV
so that FCNCs from higher-dimensional operators
are suppressed \cite{gp, hs}. Moreover, this results in small $4D$
Yukawa couplings to the Higgs,
even if there are no 
small $5D$ Yukawa couplings \cite{gp, hs}. 
The top quark can be localized near the TeV brane
to obtain a large $4D$ top Yukawa coupling.
Because the fermion profiles depend exponentially on
the bulk masses,
this provides an understanding of the
hierarchy of fermion masses (and mixing) with{\em out} hierarchies in 
fundamental ($5D$) parameters, solving
the SM flavor puzzle.

However, with bulk fermions and gauge fields, 
calculable
FCNCs from exchange of gauge Kaluza-Klein (KK) modes are induced. 
The couplings of light
fermions to gauge KK modes are non-universal which
induce FCNCs.
Unlike the flat case,
there is a significant protection from a built-in RS1-GIM~\cite{aps} due
to the approximate flatness of the KK gauge boson wavefunctions in the UV
and a hierarchy in the fermion wavefunctions in the IR;
nevertheless, the resulting contributions to FCNCs are non-negligible.
In~\cite{NMFV} it was shown that at low energies this class of
models flows to next to MFV (NMFV); that is, flavor changing effects are generated primarily through mixing with the third generation. 
Generically, within the NMFV framework flavor violation occurs 
through 
NP
 sources, with a typical scale of $\Lambda_{\rm NMFV}$, which breaks the SM flavor group
from $U(3)_Q\times U(3)_u\times U(3)_d$ down to $U(2)_Q\times U(2)_u\times U(3)_d\times 
U(1)_{\rm top}$,
where $Q,u,d$ stand for quark doublets and up and down type singlets respectively.
In addition the extra source is quasi-aligned with the SM sources of flavor breaking and the missalignment is
at most of order the CKM matrix but new sources of CP violation (CPV) are present. 
Thus, transitions between the first [second] and third generation are 
suppressed by ${\cal O} (\lambda_C^3)$  [${\cal O} (\lambda_C^2)$],
where
 $\lambda_C\sim 0.23$ is the Cabibbo mixing angle.   
Despite these suppressions,
it was recently pointed out~\cite{problem} that the presence of additional, flavor violating, right handed (RH) currents
would yield a stringent bound on 
this
framework resulting with a bound of $\Lambda_{\rm NMFV}\geq 8\,{\rm TeV}$.
This implies a rather severe little hierarchy problem.

We present a novel variant of the above models, in which at leading order (LO) flavor violation in the down type quark sector is eliminated from the theory 
and 
at
the same time leave intact the framework appealing features such as the solution of the hierarchy problem, flavor puzzle and others.
The fundamental theory is also very minimal in terms of its number of parameters and contains only four flavor violating parameters,
three mixing angles and one CPV phase.
This implies that we also eliminated the presence of other CPV, ``Majorana-like'' 
phases, which induced an RS1 CP problem~\cite{aps}.
Note that unlike in the SM, in our model the flavor violating parameters are of order unity, yet no conflict is obtained with precision flavor constraints. 

\mysection{The model}
Our set-up is very simple.
Applying the MFV paradigm~\cite{MFV} to our case we assume that 
the only sources of flavor breaking are the 5D up and down Yukawa matrices,  $Y_{u,d}$ to a bulk Higgs, $H$. 
However, unlike the 4D MFV case (or other extensions with trivial flavor structure, for example universal extra dimension~\cite{UED}) in our framework
the 5D Yukawa matrices are structureless. In other words the eigenvalues of $Y_{u,d}$  are all of the same order. Furthermore,
they are totally missaligned so that the 5D ``CKM'' matrix $V_5^{\rm KM}$ is anarchic. 

In addition, the theory contains 5D vector-like, $3\times3$, mass matrices 
$C_{Q,u,d}$
for each of the quark representations.
Bulk MFV implies that the only vector-like flavor-breaking spurions
for the doublets [singlets] are~\cite{comment} 
$Y_{u,d} Y_{u,d}^\dagger$ [$Y_{u,d}^\dagger Y_{u,d}$].  We emphasize that $V_5^{\rm KM}$  is the only source of flavor and CPV in our theory.
Under the global symmetry $U(3)_Q \times U(3)_u \times U(3)_d$, 
either $Y_u$ or $Y_d$
can be brought to diagonal form, and $V_5^{\rm KM}$ 
resides in the remaining one.  
According to our MFV assumption we can expand the 5D mass matrices as a power series in $Y_{u,d}$:
\begin{equation}
C_{u,d} = Y_{u,d}^\dagger Y_{u,d}+\dots,\,\,C_Q=  r Y_{u}Y_{u}^\dagger + Y_{d} Y_{d}^\dagger+\dots,\label{C}
\end{equation}
where universal terms and overall order one coefficients were omitted for simplicity and the dots stand for subdominant higher order terms (as discussed below).
The relevant part of the 5D Lagrangian is given by
\begin{equation} \hspace*{-.2cm}
{\cal L}_{\rm gen} =   C_{Q,u,d}
 \left(\bar{Q},\bar u,\bar d\right)\left({Q},u,d \right)
+H\ 
Y_{u,d}\bar Q\left(u,d\right), \label{model}
\end{equation}
where $C_i$ are in units of $k$ the AdS curvature,
and 
%the dimension of $Y_i$ depends
%on the details of the model, where 
we will assume that the Higgs is a 
 bulk field (see later) so that $Y_i$ are measured in units of $1/\sqrt k$.

Our first result is that despite of the fact that the fundamental theory is anarchic MFV the low energy is a hierarchic one.
This is since the eigenvalues the $C_i$ matrices are sizable, which will induce geometrical separation in the extra dimension picture or the presence of sizable anomalous dimension
in the dual conformal field theory (CFT)~\cite{rscft}.

The second, maybe less trivial result, is that this theory flows to
approximate
NMFV with additional sources  of flavor and CPV.
In order to see that recall that the 4D mass matrices for the zero modes 
can be written as~\cite{aps}
$m_{u,d}\simeq2v F_Q  Y_{u,d} F_{u,d}$, where $F_x$ correspond to the value of the quark zero-modes
on the TeV brane. 
More explicitly,  the eigenvalues $f_{x^i}$ 
of the $F_x$ matrices are given by~\cite{gp,aps}
 $f_{x^i}^2=
(1/2-c_{x^i})/( 1-\epsilon^{ 1-2c_{x^i} }
 )\,,$
where
$c_{x^i}$ are the eigenvalues of the $C_x$ matrices, $\epsilon=\exp[-\kpirc]$,
$\kpirc=\log[M_{\rm \bar Pl}/{\rm TeV}]$, $M_{\rm \bar Pl}$ is the reduced Planck mass
and $v\simeq 174{\rm GeV}$.
The 
$f_{x^i}$ correspond to the amount of compositeness of the different generations. 
The $Y_{u,d}$
are anarchic, and therefore the corresponding mixing angles are given by ratios of the $F_i$ eigenvalues.
For instance, the form of the 4D mass matrices
for the zero modes implies that the rotation to mass eigenbasis
diagonalizes 
$(m_{u,d}^2)_{ij} 
= 4v^2 (F_Q Y_{u,d} F_{u,d} F_{u,d}^\dagger Y_{u,d}^\dagger
F_Q^\dagger)_{ij} 
\sim f_{Q^i} f_{Q^j}$.  This implies that
 $(V_{\rm CKM})_{ij}\sim f_{Q^i}/f_{Q^j}$ and thus the $c_{Q^i}$ eigenvalues
control the CKM mixing angles.~\cite{aps}.

The couplings of two zero modes to the gauge KK states (which are localized near the TeV brane), have a flavor structure 
that
is different from the 4D mass matrices. They are proportional to $F_{Q,u,d}^2$,
 which is not aligned with  $m_{u,d}$.
Thus new flavor and CPV phases are present in the low energy theory. 
However,
the NMFV limit is realized since 
one eigenvalue of $(F_{u,Q,d})$ is 
much larger than the others, and thus an approximate $U(2)$ is preserved
(so that $F_{Q}^2$ and $m_{u,d}$ are quasi-aligned)~\cite{NMFV}. 
Note that the theory contains RH currents since in the mass basis the $C_{u,d}$ matrices are not diagonal.

\begin{table}[!hbt]\begin{center}
 \begin{tabular}{||c|l|l|l||}
    \hline\hline
    { Flavor}& { $c_Q,\,f_Q$} & { $c_u,\,f_u$} & { $c_d,\,f_d$}\cr
    \hline\hline
    I & 0.64,\,0.002& 0.68,\, 7\,10$^{-4}$&0.65,\,2\,10$^{-3}$\cr \hline
    II& 0.59,\,0.01& 0.53,\,0.06&0.60,\,0.008\cr\hline
    III &0.46,\, 0.2& - 0.06,\,0.8&0.58,\,0.02\cr
\hline\hline
 \end{tabular}
\caption{{\small The eigenvalues, of $C_x,F_x$ which roughly yield the right masses and CKM elements at the 
 TeV scale~\cite{hs}.}}
\end{center}\end{table}
Our third result is that in the limit where $r$ in Eq.~(\ref{C}) goes to zero, 
$C_Q$,$C_d$, and $Y_d$ can all be simultaneously diagonalized. 
 Therefore, flavor violation in the down sector is completely eliminated, where in this 
case
 flavor conversion (including the CKM part) is due to the up quark sector! 
Within our scheme and with accordance to Eq.~(\ref{C}) the value of $r$ is not a free parameter but rather a function
of the flavor parameters (which are in turn determined by the known masses and mixings).  
For concreteness, we present a numerical example that satisfies our scheme. 
In Table I we present the eigenvalues of $C_i$ and $F_i$ that yield the quark masses and mixing angles.
We further need to show that there is a consistent solution to the following relation:
\beq
{\rm diag}(C_Q) = a\, {\rm diag}[  r (V_{5}^{\rm KM\, \dagger}(\theta_{ij},\delta) C_u \,V_5^{\rm KM}(\theta_{ij},\delta)+C_d],
\label{eq:constraint}
\eeq
that is in accordance with the mass values in Table I,
where $\theta_{ij}$ is a mixing angle between the $i$th and $j$th generations and $\delta$ is the 5D CKM phase.
To see that our setup is self consistent we need to verify that the eigenvalues of the three
mass matrices, $C_{Q,u,d}$ can be derived from only two anarchical matrices, $Y_{u,d}$.
As an example the following numbers were found to solve the above relation,
$a,\,r,\,\theta_{12},\,\theta_{23},\,\theta_{13},\, \delta\,\approx 0.8,0.3,115^o,65^o,70^o,0.6\,.$

It is rather remarkable
that $r$ tends to be small. This follows
since the large top quark mass favors $c_{u^3} \ll 0.5$, 
and thus the $C_u$ eigenvalues differ in structure from the $C_Q,C_d$ 
eigenvalues (see Table I).  

To clarify this result, 
consider the constraint that Eq.~(\ref{eq:constraint})
would impose in a simpler system where only the second and third
generations are present.  Then, the $C_{Q,u,d}$ eigenvalues split up
into a trace $c_{Q,u,d}^{\rm tr} = (c_{Q^2,u^2,d^2} + c_{Q^3,u^3,d^3})/2$ 
and a traceless piece $c_{Q,u,d}^{\rm tl} = 
(c_{Q^2,u^2,d^2} - c_{Q^3,u^3,d^3})/2$. In this simpler system, Eq.~(\ref{eq:constraint}) implies only a trace condition 
$c_{Q}^{\rm tr} = ar c_u^{\rm tr} + a c_d^{\rm tr}$,
and one remaining eigenvalue condition  
$(c_Q^{\rm tl})^2 = a^2 [ r^2 (c_u^{\rm tl})^2 + (c_d^{\rm tl})^2 
+ 2 r c_u^{\rm tl} c_d^{\rm tl} \cos 2\theta_{23}]$. 
From the values in Table I, $c_u^{\rm tl} = 0.28 \gg c_Q^{\rm tl}=0.07$,
and it is straightforward to work out that these two constraints would imply
$0.19 \le |r| \le 0.31$. 
The full three-generation system, which is
necessary to obtain a CP-violating phase,
is more complicated and the allowed range of $|r|$ must be found
numerically;  
typically our numerical solutions favor $r=0.1-0.4\,.$
Thus, within our framework we find that the flavor violation in the down sector is suppressed by ${\cal O} (0.25)$ .

In our numerical examples
we have assumed that 
the typical size of the Yukawa matrix eigenvalues is $y\approx 3$
 (slightly bigger than was used in~\cite{aps} with Higgs on the brane). In theories where the Higgs is a bulk field
such as the holographic composite Higgs models~\cite{CH} $y$ is within the perturbative region for at least three KK modes, 
$N_{\rm KK}$, below the cutoff~\cite{custodial} , $N_{\rm KK} (2 y /4 \pi)^2 < 1$. 
As we shall see next, this choice yields a suppression (not due to a symmetry) of order $r_y\sim 2/3$ which together with
a moderate value of $r$ will completely relax the present tension with flavor and CPV precision bounds.

\mysection{FCNC and electric dipole moment (EDM)} 
Let us briefly review the status of the strongest constraints on the generic bulk RS1 models. 
These contributions are from  $\Delta F=2$  processes due to tree level exchange of KK gluon.
In~\cite{aps} it was shown that the ratio between the RS1, $(V-A)\times(V-A)$, contributions and the SM is proportional to 
$(F_Q^2)_{ij}^2$ (in the down quark mass basis). Using 
the relation $(V_{\rm CKM})_{ij}\sim f_{Q^i}/f_{Q^j}$ 
the ratio of contributions can be written as 
\beq
h^{\rm RS}={{M_{12}^{\rm RS}}\over M_{12}^{\rm SM}}
     \sim{0.5}\times\left({3 {\rm TeV}\over m_{\rm
           KK}}\right)^2
     \left({{ f_{Q^3} \over 0.3} }\right)^4\,. \eeq
The above contribution is proportional to $f_{Q^3}^4$ because to leading order
all flavor violation comes through the third generation.
At present, $h^{\rm RS}\lsim 0.3$~\cite{ENP,NMFV,problem}.
However, in models where RH currents are present, the dominant contributions to $\epsilon_K$ involve operators with $(V-A)\times(V+A)$ structure~\cite{problem}.
In such a case the contributions are proportional to  
$(F_Q^2)_{12} (F_d^2)_{12}\propto 
m_d m_s/(v y)^2$
 which apparently is smaller by a factor of ${\cal O} (20)$.
This is not enough due to the the following two sources of enhacement,  
${\cal O}(11)$ from chiral enhancement of the matrix element and
${\cal O}(7)$ from the running from the KK scale to the weak scale. 
These overcome the suppression and yield the largest contributions which imply that the KK masses have to be above 
the 8 TeV scale.

In our class of models both the $(V-A)\times(V-A)$ and the $(V-A)\times(V+A)$ contributions are suppressed by $r^2$. In addition, 
due to the larger overall scale for
the Yukawa matrices the value of $ f_{Q^3} $ is smaller by factor of $r_y\sim 2/3$
than in the brane-localized-Higgs case. 
Due to the RS1-GIM mechanism LH flavor violation is proportional to $ f_{Q^3} $. 
Thus these contributions are suppressed by ${\cal O}(r_y^4)$ 
where as in the case of $(V-A)\times(V+A)$ a suppressesion of ${\cal O} (r_y^2)$
is obtained. So, altogether we expect a suppression of
down quark $\Delta F=2$ currents to be of the order 
$(2/3)^{4,2} (0.25)^2 =$
${\cal O}(1,3\%)$ in the  $(V\mp A)\times(V- A)$ case, respectively.
This allows us to lower the KK masses below the 2 TeV scale without violating any of the current constrains, significantly 
below the value allowed by EW precision tests~\cite{EWT}.

Finally, we comment that (assuming a solution to the strong CP problem) our model does not suffer from a CP problem due to constraints from the neutron electric dipole moment since the contributions to this process 
arise only at two loops 
and not at one loop as occurs in the non 5D-MFV case~\cite{aps}. One way to see that
two loops are required is to compare the CPV sources of the generic case 
and our class of models.
In the general framework even in the two generation case there are various CPV phases present, so that one loop is enough in order to
be sensitive to these extra ``Majorana'' phases. However, in our case there is a single CPV phase in the fundamental theory which vanishes
in the two generations case, as the theory becomes real in that limit.
Thus only two loop diagrams can be sensitive to this 5D-CKM phase and the RS1 CP problem is solved.
In more technical terms the spurion that generates 
the leading contributions can be written, without loss of generality, as
$d_N \equiv Im\left[F_Q ( Y_u Y_u ^\dagger +Y_d Y_d ^\dagger)
 Y_d F_d\right]_{11}=  Im\left[F_Q(C_Q) ( C_Q/ar +Y_d Y_d^\dagger (1-1/r))
 Y_d F_d\right]_{11}$
 where in the RH side we have used the relation in Eq. (\ref{C}).
 This expression is only a function of $Y_d Y_d^\dagger$ and $C_Q$ where the missalignment between these two
 spurions is described by a single 5D CKM-like matrix.
 Thus for CPV all three generations must participate in the process which is possible only at two loops~\cite{Banks:1994yg}.

\mysection{Conclusions and Outlook}
We have presented a simple and economic warped extra dimension model based 
on the novel idea of 5D anarchic minimal flavor violation (MFV) in the quark sector. 
The idea carries several interesting features as follows:
The low energy theory is anarchic and the model solves the flavor puzzle; 
however
the theory is not described by MFV but rather by the next to MFV.
New flavor and CP violating phases are generically present.
However 
they dominantly induce flavor-changing currents only
in the up type sector.
In addition CP violation occurs only when three generations are considered.
Thus the agreement
with experimental constraints, both from flavor changing and flavor conserving, dipole moment experiments, is dramatically improved.
Here we focused on the quark sector.
 It would be interesting to check 
whether the above mechanism can be extended to the
lepton sector which also comes with its own flavor and CP problems~\cite{Agashe:2006iy}.
We note that no extra structure was required in order to realize the above
 scenario. Rather the number
of flavor parameters was reduced.
This implies that to a large extent the LHC collider phenomenology is similar to what was
already discussed in the context of the general framework~\cite{collider}.
We note that since flavor violation is suppressed in the down quark sector but to 
a lesser
 extent in the up 
quark
sector a possible signal
of this framework is top flavor violation~\cite{TopFCNC}. (contributions to $D-\bar D$ mixing are subdominant as in~\cite{aps})

In this work we focused on showing how our scheme solved the RS1 flavor and CP problems.
Note that the bulk flavor parameters are protected by locality since corrections to their values have to involve the two Yukawa matrices which are localized on the TeV brane.
We have not discussed how to dynamically realize the above set up but we expect that this should be rather straight forward.
One possible way is through shining~\cite{shining} from the TeV brane, where additional three bulk adjoint light scalars (transform under the quark sector flavor symmetry) 
can couple to the two TeV, bi-fundamental, Yukawa matrices. 
One can also understand/speculate from this set up why the universal contributions are somewhat suppressed
or absent since masses in the bulk must be generated by a field that is 
odd under the $Z_2$ orbifold symmetry.  
The flavons, that generate the bulk masses, are odd under the orbifold symmetry, which prevents a bare, universal, mass term.
It is not inconceivable that a solution to the strong CP problem can be also
obtained via the above setup in the spirit of~\cite{twisted}.

Our setup can be also understood from the 4D point of view where a single source of flavor breaking
induces both the mixing between the elementary and composite fermions and setting the chiral operators
anomalous dimensions but on the same time controls the structure of the purely composite
Yukawa interaction between the Higgs and fermions. 
Thus the resulting flavor violation stems from a single source.
In fact this is not completely unfamiliar since within anomaly mediation supersymmetry breaking~\cite{AM}
the flavor violation in the squark soft breaking sector is induced by the 4D Yukawa matrices.
It would be interesting to see whether a realistic supersymmetric version of the above model, 
along the lines of~\cite{NS} can be constructed.  In such a case the anomalous dimension of the operators are
proportional to the anarchic Yukawa matrices. 
The resulting flavor structure would be under better control even if 
the resulting soft masses are not degenerate.

\mysection{Acknowledgements} We thank
Kaustubh Agashe, Nima Arkani-Hamed, Ami Katz and  Martin Schmaltz for discussions.
We also thank KA for comments on the manuscript.

Note added: while this work near completion Ref.~\cite{Csaba} was published which also deals with the RS flavor problem.
However, the model of \cite{Csaba} requires
introducing the fermion mass hierarchies by hand, whereas in
our model such hierarchies are generated naturally.


\begin{thebibliography}{99}
\bibitem{comment} We assume that UV-brane localized terms 
are either universal or proportional to the Yukawa matrices as well or just small, in agreement with their naive
dimensional analysis size~\cite{NDA}. Thus our flavor structure is not spoiled by UV Planck physics.
\end{thebibliography}
\end{document}